\documentclass[11pt]{scrartcl}

\usepackage{parskip}
\usepackage[english]{babel}
\usepackage[utf8]{inputenc}
\usepackage{csquotes}
\usepackage{booktabs}
\usepackage{tikz}
\usetikzlibrary{arrows.meta}
\tikzset{>={Latex[width=3mm,length=3mm]}}
\usepackage[backend=biber, style=authoryear]{biblatex}
\usepackage[pdfpagelabels, plainpages=false]{hyperref}
\usepackage{cleveref}

\title{Resolving Aaron's Social Insurance Paradox}
\author{Martin Drees\footnote{\href{mailto:pension@mdrees.de}{pension@mdrees.de}}\\ \small{Research Institute for Discrete Mathematics, University of Bonn}\footnote{This work is independent research unrelated to the affiliation}}
\date{\today}

\tikzset{
  actor/.style={draw, rounded corners, fill=blue!10, minimum width=1.8cm, minimum height=0.8cm},
  entity/.style={draw, rounded corners, fill=green!10, minimum width=2cm, minimum height=0.8cm},
  state/.style={draw, rounded corners, fill=red!10, minimum width=2cm, minimum height=0.8cm},
  transfer/.style={->, thick, >=Latex},
  asset/.style={dashed, ->, >=Latex, draw=gray},
  flowlabel/.style={font=\small}
}

\begin{document}
\maketitle
\begin{abstract}
  This paper resolves Aaron's social insurance paradox, which suggests that introducing a pay-as-you-go (PAYG) pension system increases welfare when population growth plus average wage growth exceeds interest rates.
  Using a simplified overlapping generations model, we demonstrate this apparent advantage stems from asset reduction rather than inherent superiority.
  We analyze three pension systems—traditional PAYG, capital-funded, and capital-funded with bonus payments—and establish an equivalence between PAYG and the bonus-payment system.
  This equivalence reveals that systems with identical contributions and benefits differ only in accounting frameworks and asset positions, challenging the notion of PAYG superiority.
  Our analysis exposes a fundamental conceptual inconsistency in how sustainability is assessed across equivalent pension systems.
  As an alternative, we propose $\alpha$-stability, a framework using index shares to evaluate pension systems relative to economic indicators.
  These findings suggest that perceived advantages between pension systems often result from their formulation rather than substantive economic differences.
\end{abstract}
\paragraph{Acknowledgements}
I would like to thank Marek Góra, Meike Neuwohner and Christian Zimpelmann for their thorough feedback on an earlier draft of this paper.
I would also like to thank Henry Aaron and Hans-Werner Sinn for further helpful comments.

Additionally, I would like to thank Bert Rürup, László Végh and Christina Wilke for extensive and fruitful discussions on pension systems.

Furthermore, I would like to thank members of the Research Institute for Discrete Mathematics, the mathematical olympiad community in Germany, the QED e.V., the CdE e.V. and my family, and many others for many valuable discussions on my pension project.
\newpage

\section{Introduction}
Aaron's social insurance paradox states that \textquote{social insurance can increase the welfare of each person if the sum of the rates of growth of population and real wages exceeds the rate of interest} \parencite{Aaron1966}.

It has been described as a \textquote{path breaking article} in \textquote{The Foundations of Pension Finance} \parencite{Bodie2000}.
The paradox suggests that a pay-as-you-go (PAYG) pension system can be superior or inferior to a capital-funded system depending on whether the so-called \emph{Aaron condition} is satisfied.
Townley \parencite*{Townley1981} analyzed whether a transition between these systems can be realized in a majority voting setting.
Later, it has also been shown that there is no Pareto-improving transition from a PAYG system to a funded system even if the Aaron condition is not satisfied \parencite{Breyer1989, Fenge1995, Sinn2000, Lindbeck2003}.
In addition to model simplifications, the primary explanation for the impossibility of a Pareto-improving transition is the windfall benefits received by the first generation in a PAYG system.

In this article, we challenge Aaron's social insurance paradox on a more fundamental level.
We demonstrate that the increase in welfare can be explained by a reduction in assets.
More specifically, we present a simple example illustrating that a PAYG pension system is equivalent to a capital-funded pension system using the same amount of resources.
In particular, this equivalence shows that a PAYG pension system is not superior even if the Aaron condition is satisfied forever.

This finding more generally challenges how PAYG pension systems and funded systems are handled differently, for example in the context of sustainability.
We demonstrate that established notions of pension system sustainability are not consistent and identify a fundamental conceptual inconsistency in how sustainability is assessed across equivalent pension systems.
We explore real-world examples of this inconsistency in EU fiscal sustainability indicators \parencite{EU2022}, NDC schemes \parencite{palmer2005, Hol2017, gora2019}, and rational pension reform frameworks \parencite{Bor2007}.
To address these inconsistencies, we propose an alternative approach based on the concept of $\alpha$-stability, which extends the concept of delta-sustainability introduced in \textcite{Drees2024}.

To summarize, the goals of this paper are the following:
\begin{itemize}
\item Demonstrate the argument of Aaron's social insurance paradox in a simplified setting
\item Resolve the paradox by providing a counterexample
\item Challenge established theory and frameworks on PAYG pension systems related to Aaron's paradox:
  \begin{itemize}
  \item PAYG can be superior/inferior to capital-funded systems based on the Aaron condition
  \item The rate of return of PAYG pension systems is related to wage sum (or more generally, the contribution base)
  \item Accounting principles used for PAYG pension systems, focusing on the notion of sustainability
  \end{itemize}
\item Provide an alternative framework with a consistent analysis in the setting of Aaron's paradox
\end{itemize}

\section{Setting}
We aim to use a setting that is as simple as possible in order to clearly illustrate the core argument.

Time progresses in discrete steps, and we consider generations indexed as $0, 1, 2, \dots$.
At each time step $i$, only two generations are alive: generation $i$ (pensioners) and generation $i+1$ (workers).

Let $a_i$ denote the size of generation $i$.
We assume that both the real wage growth rate and the interest rate are constantly 0.
The population grows at a constant rate $g$.
We primarily consider the case where $g > 0$, meaning the population is growing.
Under this assumption, the Aaron condition is satisfied in our setting.

Additionally, we introduce a state that holds assets of value $A$ in the beginning.
Assets can also be negative, representing public debt.
In such cases, we assume that the interest rate on this debt is also 0.
However, the key argument does not rely on the possibility of debts.

\section{Illustrating Aaron's paradox: Three pension systems}
We now describe three different pension systems within this simple setting to illustrate the paradox.

\textbf{System P (PAYG):}
At time $i$, generation $i+1$ pays generation $i$ a total pension of $a_{i+1}$.

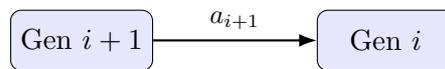
\begin{figure}[ht!]
  \centering
  \begin{tikzpicture}
    \node[actor] (P_ip1) at (0,0) {Gen $i+1$};
    \node[actor] (P_i) at (4,0) {Gen $i$};

    \draw[transfer] (P_ip1) -- node[midway, above, flowlabel] {$a_{i+1}$} (P_i);
  \end{tikzpicture}
  \caption{System P (PAYG): Pay-as-you-go pension system with direct intergenerational transfers}
\end{figure}

\textbf{System C (Capital-Funded):} At time $0$, generation $1$ purchases assets worth $a_1$.
For $i > 0$, generation $i$ sells its assets to receive a pension of $a_i$, while generation $i+1$ buys assets worth $a_{i+1}$.

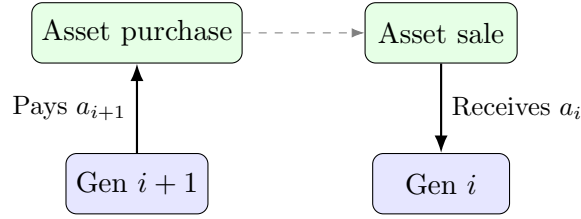
\begin{figure}[ht!]
  \centering
  \begin{tikzpicture}
    \node[actor] (C_ip1) at (0,0) {Gen $i+1$};
    \node[actor] (C_i) at (4,0) {Gen $i$};

    \node[entity] (C_asset1) at (0,2) {Asset purchase};
    \node[entity] (C_asset2) at (4,2) {Asset sale};

    \draw[transfer] (C_ip1) -- node[midway, left, flowlabel] {Pays $a_{i+1}$} (C_asset1);
    \draw[transfer] (C_asset2) -- node[midway, right, flowlabel] {Receives $a_i$} (C_i);
    \draw[asset] (C_asset1) -- (C_asset2);
  \end{tikzpicture}
  \caption{System C (Capital-Funded): Capital-funded pension system}
\end{figure}

\textbf{System CB (Capital-Funded with Bonus):}
At time $0$, the state pays generation $0$ a total pension of $a_1$, and generation $1$ purchases assets worth $a_1$.
For $i > 0$, generation $i$ sells its assets for a pension of $a_i$, while generation $i+1$ buys assets worth $a_{i+1}$.
Additionally, the state pays generation $i$ a bonus pension of $a_{i+1} - a_i$, which is positive for $g > 0$.

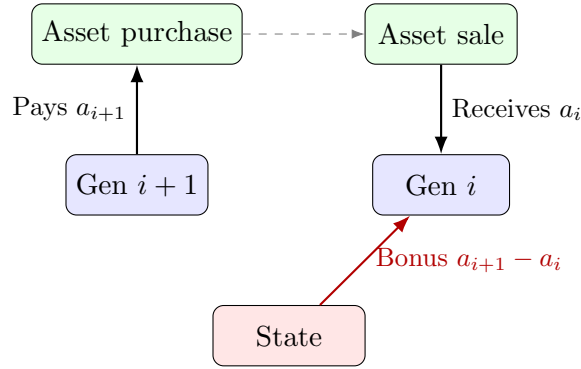
\begin{figure}[ht!]
  \centering
  \begin{tikzpicture}
    \node[actor] (CB_ip1) at (0,0) {Gen $i+1$};
    \node[actor] (CB_i) at (4,0) {Gen $i$};

    \node[entity] (CB_asset1) at (0,2) {Asset purchase};
    \node[entity] (CB_asset2) at (4,2) {Asset sale};

    \node[state] (State) at (2,-2) {State};

    \draw[transfer] (CB_ip1) -- node[midway, left, flowlabel] {Pays $a_{i+1}$} (CB_asset1);
    \draw[transfer] (CB_asset2) -- node[midway, right, flowlabel] {Receives $a_i$} (CB_i);
    \draw[asset] (CB_asset1) -- (CB_asset2);
    \draw[transfer, color=red!70!black] (State) -- node[midway, right, flowlabel] {Bonus $a_{i+1}-a_i$} (CB_i);
  \end{tikzpicture}
  \caption{System CB (Capital-Funded with Bonus): Capital-funded pension system with state bonus payments}
\end{figure}

\Cref{tab:comparison} compares the three systems for all generations and the state's assets.

\begin{table}[ht!]
  \centering
  \begin{tabular}{l l l l}
    \toprule
    & System P & System C & System CB \\
    \midrule
    Contribution of generation $i > 0$ & $a_i$ & $a_i$ & $a_i$ \\
    Benefit of generation $i > 0$ & $a_{i+1}$ & $a_i$ & $a_{i+1}$ \\
    State assets after time step $i$ & $A$ & $A$ & $A - a_{i+1}$ \\
    \bottomrule
  \end{tabular}
  \caption{Comparison of the three pension systems}\label{tab:comparison}
\end{table}

We observe that all three systems require the same contributions from all generations.
Notably, none of these systems require any contribution from generation $0$.

System P provides higher benefits than System C for every generation and additionally provides pensions for generation $0$.
Both systems keep the state's assets constant.

System CB provides the same benefits as System P by compensating for the lower benefits of System C through bonus payments.
However, this leads to a decline in state assets, which eventually become negative.

From this analysis, it appears that System P is superior to both System C and System CB:\@
\begin{itemize}
\item It provides better or at least equal benefits for every generation.
\item State assets are larger or at least equal at every point in time.
\end{itemize}

This simple example suggests an increase in social welfare when $g > 0$, even without differences in wages and real interest rates.
This increase is achieved through the introduction of a social insurance system like System P.

In Aaron's original framework, only System P and System C are introduced, albeit in a more general setting.
It seems paradoxical that reducing savings, as seen in System P, can increase the nation's overall pension benefits.

The introduction of System CB is crucial to resolve the paradox in the next section.

\section{The paradox resolved}
In this section, we resolve the paradox by addressing the following conclusion that we call \textquote{PAYG-superiority}:
\begin{quote}
  When the wage sum growth rate exceeds the interest rate, a PAYG pension system is superior to a capital-funded system.
\end{quote}

We will demonstrate that this statement is incorrect through a counterexample.
We are precisely in a setting where the PAYG-superiority would be expected to hold—with population growth (and thus wage sum growth) rate $g > 0$ exceeding the interest rate of 0.

We will show that System P and System CB are equivalent and will precisely define what this means.
Note that System P is a PAYG pension system and System CB is based on capital-funding.
The equivalence implies that System P cannot be superior to System CB, which then constitutes a counterexample to PAYG-superiority.

Afterwards, we will also compare System P and System C as in the original paradox and argue that PAYG-superiority is misleading in this comparison.

\subsection{Comparison of worker and state assets}

Before explaining the concept of equivalence, we first consider the distribution of assets in both systems at different time points, as shown in \Cref{tab:asset_comparison}.

\begin{table}[ht!]
  \centering
  \begin{tabular}{l l l}
    \toprule
    & System P & System CB \\
    \midrule
    Worker assets after step $i$ & $0$ & $a_{i+1}$ \\
    State assets after step $i$ & $A$ & $A - a_{i+1}$ \\
    \bottomrule
  \end{tabular}
  \caption{Comparison of worker and state assets in System P and System CB}\label{tab:asset_comparison}
\end{table}

While these systems appear different in their asset distributions, an important observation is that total assets are equal.

\subsection{Equivalence by transition}

We will show that System P and System CB are equivalent in the following sense.
At every point in time $i$, one can transition from one system to the other and reach the same status as one would have reached by running the other system from the beginning.
Here, the status is defined by the assets of the workers and the state.
The transition does not change any benefits or contributions.

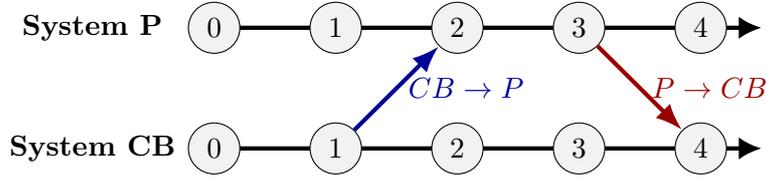
\begin{figure}[ht!]
  \centering
  \usetikzlibrary{positioning, calc, decorations.pathreplacing}
  \tikzset{
    timenode/.style={draw, circle, minimum size=0.6cm, fill=gray!10},
    systemlabel/.style={font=\bfseries},
    transform/.style={->, ultra thick, >=Latex},
    timeline/.style={ultra thick, -}
  }
  \begin{tikzpicture}[scale=0.8]

    \draw[timeline, ->] (0,1) -- (9,1);
    \draw[timeline, ->] (0,-1) -- (9,-1);

    \node[systemlabel] at (-2, 1) {System P};
    \node[systemlabel] at (-2, -1) {System CB};

    \foreach \x in {0,1,...,4} {
      \node[timenode] (P\x) at (2*\x, 1) {\x};
      \node[timenode] (CB\x) at (2*\x, -1) {\x};
    }

    \draw[transform, blue!60!black] (CB1) to node[midway, right] {$CB \to P$} (P2);
    \draw[transform, red!60!black] (P3) to node[midway, right] {$P \to CB$} (CB4);

  \end{tikzpicture}
  \caption{Equivalence of System P and System CB by possible transitions between the systems}
\end{figure}

In particular, this equivalence shows that System P is not superior to System CB, because each generation can decide to switch between the two systems.
We now describe the specific transitions in detail:

\textbf{Transition $P \to CB$ at time $i$:} Generation $i$ receives its benefits from the state, reducing state assets to $A - a_{i+1}$.
Additionally, generation $i+1$ purchases assets worth $a_{i+1}$.

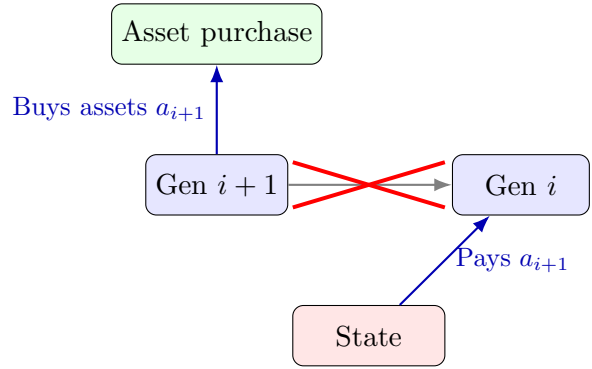
\begin{figure}[ht!]
  \centering
  \begin{tikzpicture}
    \node[actor] (P_ip1) at (0,0) {Gen $i+1$};
    \node[actor] (P_i) at (4,0) {Gen $i$};
    \node[entity] (C_asset1) at (0,2) {Asset purchase};
    \node[state] (State) at (2,-2) {State};

    \draw[transfer, color=gray] (P_ip1) -- (P_i);
    \draw[color=red, line width=1.5pt, -] (1,0.3) -- (3,-0.3);
    \draw[color=red, line width=1.5pt, -] (1,-0.3) -- (3,0.3);

    \draw[transfer, color=blue!70!black] (P_ip1) -- node[midway, left, flowlabel] {Buys assets $a_{i+1}$} (C_asset1);
    \draw[transfer, color=blue!70!black] (State) -- node[midway, right, flowlabel] {Pays $a_{i+1}$} (P_i);
  \end{tikzpicture}
  \caption{Transition of System P to System CB at time $i$}
\end{figure}

\textbf{Transition $CB \to P$ at time $i$:}
Generation $i$ sells its assets and receives the bonus as in the continuation of System CB.\@
This sets state assets to $A - a_{i+1}$ after the bonus payment.
Generation $i+1$ then pays the state $a_{i+1}$ instead of buying assets.
This will adjust the state assets back to $A$, and generation $i+1$ will have no assets, as in System P.

\begin{figure}[ht!]
  \centering
  \begin{tikzpicture}
    \node[actor] (P_ip1) at (0,0) {Gen $i+1$};
    \node[actor] (P_i) at (4,0) {Gen $i$};
    \node[entity] (CB_asset1) at (0,2) {Asset purchase};
    \node[entity] (C_asset2) at (4,2) {Asset sale};
    \node[state] (State) at (2,-2) {State};

    \draw[transfer, color=gray] (P_ip1) -- node[midway, left, flowlabel] {Buys assets} (CB_asset1);
    \draw[transfer] (CB_asset2) -- node[midway, right, flowlabel] {Receives $a_i$} (CB_i);
    \draw[transfer, color=red!70!black] (State) -- node[midway, right, flowlabel] {Bonus $a_{i+1}-a_i$} (CB_i);

    \draw[color=red, line width=1.5pt, -] (-0.4,0.8) -- (0.4,1.2);
    \draw[color=red, line width=1.5pt, -] (-0.4,1.2) -- (0.4,0.8);

    \draw[transfer, color=blue!70!black] (P_ip1) -- node[midway, left, flowlabel] {Pays $a_{i+1}$} (State);
  \end{tikzpicture}
  \caption{Transition of System CB to System P at time $i$}
\end{figure}
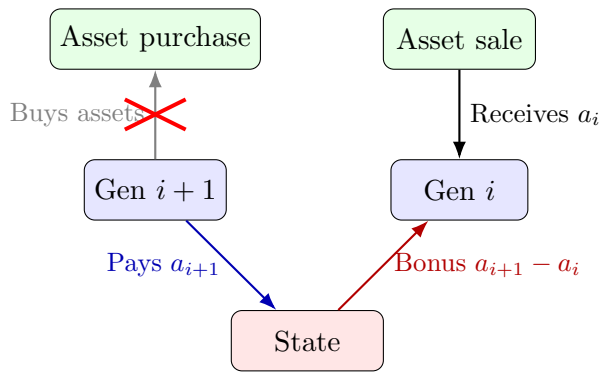

\subsection{Implications of equivalence}

The equivalence demonstrated above also has profound implications for a comparison between System P and System C similar to the original setting of Aaron's paradox.

Rather than comparing System P and System C, we can now compare System C and System CB, which provides a clearer perspective.
In this comparison, the statement that System CB increases welfare is misleading, as this increase depends entirely on using additional state assets.

From this perspective, the resolution of Aaron's paradox becomes clear: The apparent advantage of PAYG systems when the wage sum growth rate exceeds the interest rate is not due to an inherent efficiency property of such systems, but rather a consequence of accounting frameworks that obscure the true cost in terms of state asset reduction or implicit debt.

\subsection{Addressing potential objections}

The equivalence between System P and System CB presented above raises several potential objections that we must address to validate our counterexample to PAYG-superiority.

\paragraph{The role of the state}
The equivalence between System P and System CB depends on assuming a state with non-constant assets.
However, Aaron's original argument did not involve a state, so we must address whether this assumption undermines the argument or distorts the comparison.

Although the existence of a state is not inherently a strong assumption, one could argue that System CB relies on the availability of assets, which may not always be feasible.
In this case, the equivalence between systems may not be relevant, as System CB could fail to function without sufficient assets to support it.

In settings where public debt is not possible, a PAYG pension system indeed offers an advantage: it effectively functions as a mechanism for taking on debt.
This setting is also the one initially explored by \textcite{Samuelson1958}.

However, if we allow for public debt and assume that the interest rate on public debt is equal to the interest rate on assets, then the equivalence between System P and System CB becomes fully operational.

\paragraph{Impact of debt costs on system choice}
In a situation where public debt is necessary but the cost of debt exceeds the interest rate on assets, the dynamics of the system change.
Under these conditions, maintaining System CB is less attractive.

System P offers an advantage in this case.
By avoiding the need to maintain both assets and debt simultaneously, System P is a more efficient choice when the state faces the challenge of high-interest debt.

\paragraph{Distinction from Aaron's paradox}
While the argument above illustrates the advantage of System P when debt costs are high, it diverges from the core insight of Aaron's paradox.
The central idea of Aaron's paradox is that the rate of return on System P is fundamentally influenced by population growth, which has a direct impact on the contributions and benefits in the system.

In Aaron's framework, the key insight is that if the growth rate of population and real wages exceeds the interest rate, social welfare can be increased by introducing a pay-as-you-go (PAYG) system like System P.
The effect of population growth leads to increasing contributions over time.

In contrast, the argument about debt costs and the need for assets focuses on fiscal considerations unrelated to population growth.
While System P's avoidance of both debt and asset management provides advantages in the face of high debt costs, this is a different issue from the demographic forces at play in Aaron's paradox.

\paragraph{System P vs. System C without state and no debt}
In the scenario where there is no state and no debt, we return to the original framework of Aaron's paradox, considering only System P and System C.
An important observation is that, at any point in time, the population could claim the assets of the workers and transition to a pay-as-you-go system like System P.

The critical point in this context is that System P appears superior to System C because the additional assets in System C are never actively used.
These assets are held indefinitely, and because they are not put to any active purpose or liquidation, they effectively have no value.
While this observation holds theoretically, it is not particularly relevant in practical terms.
In real-world scenarios, assets are typically used or eventually liquidated, so the distinction between the two systems becomes less significant outside the context of this simplified theoretical framework.

\section{Inconsistent and misleading terminology}

In this section, we explore how the equivalence of System P and System CB can help clarify and challenge common terminology used in the analysis of pension systems.
A term or concept is considered inconsistent if it leads to different assessments of the same situation under different formulations, depending on how the term is applied.

\subsection{Rate of return}
When comparing pension systems, the concept of rate of return is often invoked.
For System P, the natural answer is that the rate of return equals the growth rate $g$ of the population.
For System C, the rate of return is simply 0, reflecting the zero interest rate in our setting.
For System CB, the answer is ambiguous—individuals experience a rate of return similar to System P, while the systemic rate could be considered 0.

This inconsistency reveals that the concept of rate of return can create the misleading impression that one system inherently produces higher returns than another.
However, our analysis shows that these systems are fundamentally equivalent at the systemic level.
Therefore, we suggest that the term \textquote{rate of return} should be used with caution when analyzing pay-as-you-go pension systems.

\subsection{Sustainability}
Consider the term \textquote{sustainability} as it is applied to pension systems and the following two contrasting statements:

\begin{itemize}
\item \textbf{System P is sustainable.} This is because it operates on a pay-as-you-go basis, where each generation contributes to the pension of the previous generation, without requiring the accumulation or management of assets.
  System P functions indefinitely, with the ongoing intergenerational transfers of resources.
\item \textbf{System CB is not sustainable.} This statement reflects that System CB relies on assets and the eventual consumption of these assets, which could be seen as unsustainable.
\end{itemize}

However, despite these differing assessments, \textbf{System P and System CB are equivalent} in the sense that one can transition from one system to the other at any point in time without altering the future contributions and benefits.
The apparent contradiction is resolved with the realization that the systems are fundamentally the same, and the distinction in sustainability comes down to different interpretations of what makes a system \textquote{sustainable}.

\paragraph{Extended systems: $P_\gamma$ and $CB_\gamma$}
To further elaborate on the concept of sustainability and clarify the implications of different systems, we introduce a more general setting.
Let us consider the systems $P_\gamma$ and $CB_\gamma$, where every contribution and benefit at time $i$ is multiplied by a constant factor $ \gamma $.
For instance, in the $P_\gamma$ system, generation $i+1$ would pay generation $i$ a pension of $ \gamma \cdot a_{i+1} $.
Similarly, the $CB_\gamma$ system would also scale all contributions and benefits by $ \gamma $.
Both systems would remain equivalent for all values of $ \gamma $ since we can still apply the same transition between $P_\gamma$ and $CB_\gamma$ at any point in time.

\begin{table}[ht!]
  \centering
  \begin{tabular}{l l l}
    \toprule
    & $P_\gamma$ & $CB_\gamma$ \\
    \midrule
    Contribution of generation $i>0$ & $ \gamma \cdot a_i $ & $ \gamma \cdot a_i $ \\
    Benefit of generation $i$ & $ \gamma \cdot a_{i+1} $ & $ \gamma \cdot a_{i+1} $ \\
    State assets after time step $i$ & $A$ & $A - \gamma \cdot a_{i+1}$ \\
    \bottomrule
  \end{tabular}
  \caption{Comparison of $P_\gamma$ and $CB_\gamma$ for different values of $ \gamma $.}
\end{table}

When comparing two versions of the PAYG system, say $P_1$ and $P_2$, and declaring both of them to be \textquote{sustainable}, we are missing an important distinction: the systems differ in their levels of asset accumulation.
While $P_1$ and $P_2$ both cover benefits from contributions, the amounts of assets held in $CB_1$ and $CB_2$ clearly differ.
The state in $CB_1$ holds more assets than in $CB_2$ at the same point in time.
This reveals that labeling both $P_1$ and $P_2$ as \textquote{sustainable} does not capture the reality of the state's asset situation.

Sustainability is, therefore, not a suitable term when analyzing the state's situation as a whole.
It can mislead us into thinking that a pension system labeled as \textquote{sustainable} can be treated similarly to a funded system with no debt, ignoring the fact that the state's asset position may differ in an equivalent formulation depending on the system and the value of $ \gamma $.

It may be more useful to describe System P as \emph{permanently liquid}, emphasizing its operation without the need for asset accumulation.

We emphasize once more that transitioning System P into System CB does not affect any current or future contributions or benefits.
It represents a change in the system's formulation rather than a substantive systemic reform.

\section{An established conceptual inconsistency}

In this section, we identify a fundamental inconsistency in how sustainability is conceptualized across different pension systems.
By examining how equivalent systems respond differently to the same scenario under conventional sustainability metrics, we reveal inherent contradictions in these metrics.

\subsection{A revealing scenario}

Consider a point in time $i$ and imagine increasing all contributions and benefits at and after that time by a factor of $\beta > 1$.
Specifically, all future contributions and benefits from time $i$ onward are scaled by this factor $\beta$.

In System P, this scenario would immediately generate surpluses at time $i$.
Generation $i+1$ would contribute $\beta \cdot a_{i+1}$ while generation $i$ would still receive only $a_{i+1}$, creating a surplus of $(\beta-1) \cdot a_{i+1}$.
In our setting, this surplus could be used to increase state assets.

In System CB, the scaling by $\beta$ affects both the amount of assets purchased by generation $i+1$ (which increases to $\beta \cdot a_{i+1}$) and the state bonus payments for future time periods $j > i$ (which increase to $\beta \cdot (a_{j+1} - a_j)$).
The net effect on state assets is negative due to the increased bonus payments in future periods.

When viewed through traditional sustainability frameworks, this scenario appears positive for System P (increasing state assets) but negative for System CB (increasing state expenditure).
However, this assessment is conceptually inconsistent, as we have demonstrated that System P and System CB are fundamentally equivalent.
The same policy change should not yield contradictory sustainability evaluations across equivalent systems.

\subsection{Real-world examples of conceptual inconsistency}

The inconsistency identified above is not merely theoretical but manifests in several widely-used sustainability concepts in pension policy.
Importantly, most of these concepts focus only on the pension system in isolation, without considering the broader fiscal framework.
When analyzing the pension system alone, the scenario appears positive, while a more comprehensive analysis of System CB reveals that the state faces increased bonus payments, which is negative from a fiscal perspective.

\paragraph{EU sustainability indicators}
The fiscal sustainability indicators S1 and S2 used in the EU \parencite{EU2022} exemplify this inconsistency.
These indicators consider explicit debts and their associated interest rates alongside future primary balances.

For System P, the effects on primary balances are consistently zero, as contributions equal pension expenses in each period.
For System CB, however, the effects on primary balances are negative due to the bonus payments.
Since the initial debt is identical in both systems, these fiscal sustainability indicators would yield inconsistent assessments despite the systems' equivalence.

\paragraph{NDC and financial balance}
The Nonfinancial Defined Contributions (NDC) scheme, as described in detail by \textcite{palmer2005} and introduced non-technically by \textcite{Hol2017}, employs nonfinancial accounts where contributions are recorded but not invested in real assets.\ \textcite{gora2019} describe NDC as the generic old-age pension scheme.
This scheme uses a concept of financial balance that compares assets and liabilities, where liabilities correspond to existing pension claims and assets are based on future contributions.

In our scenario of scaling by $\beta$, the NDC approach would show increased assets (future contributions) without directly changing the existing liabilities (pension claims).
The system would definitely improve its financial balance, suggesting a positive effect on sustainability.
However, this fails to capture the increased expenditure in System CB due to larger state bonus payments in the future, demonstrating another inconsistency in sustainability assessment.

\paragraph{Rational pension reform framework}
\textcite{Bor2007} proposes a framework for what he terms \textquote{rational pension reform} in PAYG systems, where contributions equal benefits.
This framework, which is more general than System P in our simplified setting, would also display inconsistencies in our scaling scenario.

In this framework, the increased contributions at time $i$ would allow for an immediate increase in benefits to the pension generation, appearing to generate a \textquote{free lunch}.
While this does not directly benefit sustainability, the scaling could be a means to replace additional payments that would otherwise be provided by the state directly.
The apparent improvement stems from the framework's definitional constraints rather than genuine economic gains.

These examples demonstrate that well-established sustainability concepts can yield inconsistent assessments when applied to equivalent pension systems.
This inconsistency undermines the reliability of these concepts as guides for pension policy and suggests the need for alternative approaches.

\section{An alternative approach for pension system sustainability}

Having identified inconsistencies in conventional sustainability concepts, we now propose an alternative approach that provides a more coherent framework for assessing pension system sustainability.

\subsection{Applying the index shares concept}

Building upon the framework introduced in \textcite{Drees2024}, we apply the concept of index shares to our simplified setting of System P and System CB.\@
Note that the concept is renamed from index points to index shares to avoid confusion with point-based pension systems.

Index shares represent standardized units of pension claims, where each unit corresponds to a claim on future resources.
An index share is defined by its relationship to an economic indicator: one index share has the value of the selected economic indicator at every point in time.
This creates a direct link between the pension system and the broader economy.
Index shares can be held by participants or by the state, and their ownership represents claims on future resources.
They act as an accounting device that links contributions to benefits, with their value changing according to the chosen economic indicator over time.
This concept of index shares is very similar to nonfinancial accounts in NDC systems, although with a more explicit connection to economic indicators.
To avoid dealing with very small fractions of index shares, one can introduce a unit whose value is a constant fraction of an index share.

We choose working population growth (equivalent to population growth and wage sum growth in our setting) as the indexation method, such that the value of an index share at time $i$ is $a_{i+1}$.

In System P, generation $i+1$ buys one index share from generation $i$ at time $i$.
Generation $i$ receives $a_{i+1}$ for this index share, reflecting its current value.
The transaction is direct, without state involvement, and the total amount of index shares in circulation remains constant at one.

In System CB, the state holds the index share.
The increase in the value of this index share from $a_i$ to $a_{i+1}$ at each time step is used to finance the bonus payments of $a_{i+1} - a_i$ to generation $i$.

The key insight is that the total amount of index shares represents a global deficit (\textquote{pension debt}) that remains constant and equal across both systems.
When the state holds index shares, these can be interpreted as assets offsetting this debt.

After paying generation $0$ its initial pension in System CB, the total assets of the state are $A - a_1 + a_1 = A$, which is identical to its assets in System P.
This reflects the fundamental equivalence of the two systems.

Furthermore, in the extended systems $P_\gamma$ and $CB_\gamma$ introduced earlier, there would be $\gamma$ index shares in circulation, with proportionally scaled contributions and benefits.

\subsection{Delta-sustainability and its shortcomings}

The concept of delta-sustainability, as defined in \textcite{Drees2024}, states that a pension system is delta-sustainable if the sum of unfunded liabilities relative to an economic indicator does not increase over time, and any transfer of value into the system decreases unfunded liabilities by that amount.

While this concept addresses some inconsistencies, it has several shortcomings:

First, while state transfers are accounted for at the time of transfer, delta-sustainability does not fully consider what happens after the transfer.
One needs to account for the relationship between the increase in pension claims (e.g., in index shares) and interest rates on public debt over time.

Second, delta-sustainability does not adequately reflect the current status of a pension system.
Different initial allocations of index shares are not assessed differently, despite potentially significant variations in sustainability implications.

\subsection{\texorpdfstring{$\alpha$-stability}{alpha-stability}: A proposal to address shortcomings}

To address these limitations, we introduce the concept of $\alpha$-stability relative to an economic indicator.

A pension system is $\alpha$-stable relative to an economic indicator if it operates with a constant amount of $\alpha$ index shares, where the value of an index share is equal to the economic indicator.

In our framework, System P and System CB are $1$-stable relative to working population size.
In the extended systems $P_\gamma$ and $CB_\gamma$ introduced earlier, there would be $\gamma$ index shares in circulation, making these systems $\gamma$-stable relative to working population size.

This concept provides several advantages:
\begin{itemize}
\item It captures both the current state of the system (through the value of $\alpha$) and its evolution over time (through the stability requirement and the changing value of the economic indicator).
\item It allows for consistent comparison across different pension systems by standardizing their assessment through index shares.
\item It integrates the pension system with broader fiscal considerations by relating index shares to economic indicators.
\item It avoids the inconsistencies identified in traditional sustainability metrics by focusing on the fundamental equivalence of seemingly different systems.
\end{itemize}

\subsection{Aaron's paradox through the lens of index shares}

The original formulation of Aaron's paradox, which compares System P and System C, can be illuminated through the index shares framework.
System P effectively operates with one index share circulating through the economy.
This index share is held by generation $i$ and sold to generation $i+1$ at each time step.
The value of this index share grows with the population, providing the welfare increase described in Aaron's paradox.

In contrast, System C operates with zero index shares from a system perspective.
Each generation creates its own temporary assets that are not systemically interconnected across time periods in the same way.
There is no mechanism for value to grow with the population as in System P.

This difference in the number of index shares—one versus zero—explains why System P appears superior when the Aaron condition is satisfied.
The presence of an index share in System P allows the system to capture the growth in population, which System C cannot do.
However, this is not due to an inherent economic advantage but rather to the formulation of the system and its accounting framework.

The perspective of $\alpha$-stability makes this distinction clear: System P is $1$-stable relative to population size, while System C is $0$-stable.
The apparent superiority of System P stems from this difference in stability parameters rather than from any fundamental economic advantage.

\subsection{Comparison with existing sustainability notions}

The concept of $\alpha$-stability is conceptually similar to delta-sustainability, as both focus on the evolution of unfunded liabilities relative to economic indicators.
However, $\alpha$-stability addresses the shortcomings identified in delta-sustainability.

Various notions of pension system sustainability exist in the literature, each offering a unique perspective.
As discussed in \textcite{Drees2024}, traditional approaches to measuring pension system sustainability, such as those outlined by \textcite{holzmann2004} and \textcite{devesa2010}, focus on either implicit debt in isolation or actuarial balance within closed systems.

\textcite{holzmann2004} emphasize the importance of measuring implicit debt and suggest assessing pension reforms by considering explicit debt and implicit debt jointly, but do not provide an explicit tool to do so.
The notion of $\alpha$-stability constitutes a way to implement this suggestion.

\textcite{devesa2010} note that solely measuring implicit debt does not provide a comprehensive definition of pension system sustainability, as it remains unclear what the maximum level of implicit debt should be.
To address this, they introduce the concepts of actuarial imbalance and unitary pension cost, comparing the total benefits received by a group of participants to their total contributions.
This idea resembles the notion of $\alpha$-stability in that the sustainability of the running system is considered.
Nevertheless, their approach assumes a closed system and does not directly account for changes in unfunded liabilities.

The $\alpha$-stability concept, like delta-sustainability, recognizes that the pension system is not a closed entity but is part of the state's overall fiscal position.
It acknowledges that sustainability should be evaluated in terms of the system's impact on the state's overall financial position rather than in isolation.

The $\alpha$-stability concept avoids the inconsistencies inherent in traditional metrics when applied to equivalent pension systems.

\section{Conclusion}

Aaron's social insurance paradox suggests that social insurance can increase welfare when the sum of the growth rates of population and real wages exceeds the rate of interest.
In this paper, we have demonstrated that this increase in welfare can be understood as a result of reduced asset accumulation rather than a superior system.

We have shown that a pay-as-you-go (PAYG) pension system (System P) is fundamentally equivalent to a capital-funded system when resources are consumed in the same way (System CB).
This challenges the natural impression that one system is inherently superior.

This equivalence has far-reaching implications beyond resolving Aaron's paradox.
It reveals a conceptual inconsistency in how sustainability is assessed across different pension systems.
We demonstrated that conventional sustainability metrics yield contradictory evaluations when applied to equivalent systems.
This inconsistency manifests in various real-world contexts, including EU fiscal sustainability indicators, NDC frameworks, and rational pension reform theories.

To address these inconsistencies, we developed an alternative approach based on the concept of $\alpha$-stability.
This framework uses index shares to provide a standardized method for evaluating pension systems.
By relating these index shares to economic indicators, $\alpha$-stability captures both the current state of a pension system and its evolution over time.
Unlike traditional approaches, it acknowledges that pension systems are not closed entities but integral parts of broader fiscal frameworks.

Our analysis allows us to reassess and clarify common notions and concepts, encouraging a more consistent understanding of pension systems and their underlying mechanics.
This improved conceptual clarity can guide more coherent pension policy development and reform.

\paragraph{AI Usage Declaration}
This article was developed with the assistance of AI language models, specifically Claude 3.7 Sonnet and ChatGPT (models GPT-4o and o1), for tasks including initial formulation, grammar and syntax refinement, formatting suggestions, and LaTeX editing.
The underlying research, mathematical analysis, concepts, and arguments are the original contribution of the author.
The author takes full responsibility for the content, accuracy, and conclusions presented in this article.

\printbibliography{}
\end{document}